\documentclass[superscriptaddress, prb, twocolumn]{revtex4}

\usepackage{graphicx}% Include figure files
\usepackage{verbatim}
\usepackage{amsmath,amssymb,relsize}

\makeatletter
\def\Ddots{\mathinner{\mkern1mu\raise\p@
\vbox{\kern7\p@\hbox{.}}\mkern2mu
\raise4\p@\hbox{.}\mkern2mu\raise7\p@\hbox{.}\mkern1mu}}
\makeatother
\usepackage{amsmath}
\usepackage{amsthm}
\usepackage{amssymb}
\usepackage{rotating}
\usepackage{epstopdf}
\usepackage{bbold}
\usepackage{times}
\usepackage[none]{hyphenat}
\usepackage{threeparttable}

\def\Inf{\operatornamewithlimits{inf\vphantom{p}}}

\usepackage{color}
\usepackage{mathrsfs}

\usepackage[normalem]{ulem}

\begin{document}

\author{Giuseppe Pica}
\affiliation{SUPA, School of Physics and Astronomy, University of St Andrews, KY16 9SS, United Kingdom}
\author{Gary Wolfowicz}
\affiliation{London Centre for Nanotechnology, University College London, London WC1H 0AH, United Kingdom}
\affiliation{Dept. of Materials, Oxford University, Oxford OX1 3PH, United Kingdom}
\author{Matias Urdampilleta}
\affiliation{London Centre for Nanotechnology, University College London, London WC1H 0AH, United Kingdom}
\author{Mike L. W. Thewalt}
\affiliation{Dept. of Physics, Simon Fraser University, Burnaby, British Columbia V5A 1S6, Canada}
\author{Helge Riemann}
\affiliation{Institute for Crystal Growth, Max-Born Strasse 2, D-12489 Berlin, Germany}
\author{Nikolai V. Abrosimov}
\affiliation{Institute for Crystal Growth, Max-Born Strasse 2, D-12489 Berlin, Germany}
\author{Peter Becker}
\affiliation{Physikalisch-Tecnische Bundesanstalt, D-38116 Braunschweig, Germany}
\author{Hans-Joachim Pohl}
\affiliation{Vitcon Projectconsult GmbH, 07745 Jena, Germany}
\author{John J. L. Morton}
\affiliation{London Centre for Nanotechnology, University College London, London WC1H 0AH, United Kingdom}
\affiliation{Dept. of Electronic $\mathscr{E}$ Electrical Engineering, University College London, London WC1E 7JE, United Kingdom}
\author{R. N. Bhatt}
\affiliation{Dept. of Electrical Engineering, Princeton University, Princeton, New Jersey 08544, USA}
\author{S. A. Lyon}
\affiliation{Dept. of Electrical Engineering, Princeton University, Princeton, New Jersey 08544, USA}
\author{Brendon W. Lovett}
\affiliation{SUPA, School of Physics and Astronomy, University of St Andrews, KY16 9SS, United Kingdom}
\affiliation{Dept. of Materials, Oxford University, Oxford OX1 3PH, United Kingdom}

\title{Hyperfine Stark effect of shallow donors in silicon}

\begin{abstract}
We present a complete theoretical treatment of Stark effects in doped silicon, whose predictions are supported by experimental measurements. A multi-valley effective mass theory, dealing non-perturbatively with valley-orbit interactions induced by a donor-dependent central cell potential, allows us to obtain a very reliable picture of the donor wave function within a relatively simple framework. Variational optimization of the $1s$ donor binding energies calculated with a new trial wave function, in a pseudopotential with two fitting parameters, allows an accurate match of the experimentally determined donor energy levels, while the correct limiting behavior for the electronic density, both close to and far from each impurity nucleus, is captured by fitting the measured contact hyperfine coupling between the donor nuclear and electron spin.

We go on to include an external uniform electric field in order to model Stark physics: With no extra \emph{ad hoc} parameters, variational minimization of the complete donor ground energy allows a quantitative description of the field-induced reduction of electronic density at each impurity nucleus. Detailed comparisons with experimental values for the shifts of the contact hyperfine coupling reveal very close agreement for all the donors measured (P, As, Sb and Bi). Finally, we estimate field ionization thresholds for the donor ground states, thus setting upper limits to the gate manipulation times for single qubit operations in Kane-like architectures: the Si:Bi system is shown to allow for $A$ gates as fast as $\approx$10 MHz.
\end{abstract}

\maketitle

\section{Introduction} Donor spins in silicon represent one of the most promising and well studied candidates for quantum computing architectures~\cite{RevModPhys.85.961}. Very long coherence times have been measured in both nuclear \cite{saeedi13} and electron spin donor qubits \cite{tyryshkin12}, and individual spins can be manipulated and measured~\cite{morello10,pla13,pla12,spinmem}. In any large scale information processing architecture, the application of an electric field is likely to be a vital enabling tool for addressing individual qubits~\cite{kane98}.
Whatever the specific setting used, manipulation of quantum information in these systems requires a thorough understanding of how the energy levels of the spin qubits are modified by external magnetic or electric fields. These could either be deliberately applied to execute a particular gate operation, or exist anyway in an inhomogeneous electrostatic environment. Electric fields in particular can strongly affect two main properties of the donor via the Stark effect: the hyperfine coupling between the nuclear and the electron spin, proportional to the electronic density at the nuclear site, and the electron $g$-factor, i.e. the splitting induced by a magnetic field between the spin up and the spin down electronic levels.

Knowledge of such effects is ever more critical when an electric field is used to directly engineer the electronic wave function for storage or manipulation of quantum states~\cite{RevModPhys.85.961,kane98,proposal,lansbergen2008,wave functioncontrol}. The first and most famous proposal of this kind was provided by Kane \cite{kane98}, where a single qubit state, encoded in the impurity nuclear spin, is manipulated by using a Stark shift to bring it into resonance with an oscillating magnetic field. More recently, scalable architectures have been proposed to extend single qubit control techniques to larger structures \cite{buch13,twogatehyp}.

For these reasons, the Stark effect in doped silicon has been broadly studied in literature, either theoretically \cite{tb,PhysRevLett.94.186403,kaneoperations,PhysRevLett.102.017603,Debernardi2006,PhysRevB.88.045307,PhysRevB.80.165314,PhysRevLett.99.036403} or experimentally \cite{PhysRevLett.97.176404,arsenicstark,PhysRevB.1.4071,PhysRev.124.1068}. More generally, the ability to theoretically describe the donor electron wave function accurately in a wide range of electrostatic environments is beneficial for determining the values of control parameters which provide best performance, and in the best case for estimating \emph{a priori} the feasibility of quantum algorithms and error correction codes \cite{PhysRevB.80.165314}. 

Nonetheless, the physical mechanisms underlying Stark effect of donors in silicon are not yet fully understood. 
The best attempt so far within effective mass theory (EMT) was proposed by Friesen \cite{PhysRevLett.94.186403}:  For Si:P he correctly predicted a quadratic hyperfine shift, but one which is one order of magnitude larger than the value expected from the two measurements performed so far (Si:Sb \cite{PhysRevLett.97.176404} and Si:As \cite{arsenicstark}). More recently, other theories such as tight binding (TB) and Band Minima Basis (BMB) \cite{PhysRevLett.99.036403} have been applied to the same Si:P problem, leading to closer agreement with experiment. TB and BMB, though, are computationally demanding numerical approaches, which eclipse full physical understanding. Current theoretical predictions of hyperfine shifts, then, are limited to Si:P alone, but other V group donors such as Bi are now established as promising alternative donor qubits and are being widely researched~\cite{PhysRevLett.105.067601,clocktransitions,PhysRevLett.105.067602,activation,natureactivation}. We are thus motivated to present a multi-valley EMT that provides a unifying framework for P, As, Sb and Bi donors. Theoretical predictions are supported by complete and precise experimental work.

%%the sign of such curvature is predicted to depend on the relative angle between electric and magnetic fields %%\cite{PhysRevB.80.155301}. Previous experimental investigations \cite{PhysRevLett.97.176404}, though, displayed opposite %%signs to those predicted by theory. 
Finally, we will consider the effects of the electric field on the donor ground binding energy. This has raised opposing opinions \cite{PhysRevLett.94.186403,Debernardi2006} as which of two competing effects play a dominant role in determining its magnitude: the lowering of single-valley energies due to admixing higher orbital states in the ground level, \emph{vs} the narrowing of the valley-orbit $1s$ spectrum, which is a consequence of the reduced effect of the short-range impurity potential on the energy levels when the electron moves away from the nucleus. We clarify how the interplay of both results in an overall energy decrease of the ground donor state under the external field, confirming earlier \emph{ab initio} calculations \cite{Debernardi2006,PhysRevB.80.165314}. 

The highly successful match between our theory and experimental measurements motivates the determination of expected field ionization thresholds for each implanted donor species, setting $i)$ upper limits to the achievable speeds for single-qubit operations relying on resonant excitation of selected donor electron spin transitions (like the Kane architecture \cite{kane98}); and $ii)$ gate voltages which should be applied to read-out the state of bulk qubits \cite{pla12,pla13}, possibly following transfer of quantum information to the electron spin from other degrees of freedom \cite{proposal}. 

\section{Theory}
The Hamiltonian of a donor electron weakly bound to an impurity nucleus implanted in a silicon lattice is:
\begin{equation}\label{eqbase}
H \Psi(\textbf{r})= \left[-\frac{\hbar^{2}}{2 m_{0}}\nabla^{2} + V^{0}(\textbf{r}) + U(\textbf{r})+e \textbf{E} \cdot \textbf{r} \right]\Psi(\textbf{r})= \epsilon \Psi(\textbf{r}),
\end{equation}   
where $\Psi(\textbf{r})$ is the wave function of the donor electron, $m_{0}$ is its rest mass, $V^{0}(\textbf{r})$ is the periodic potential of the undoped silicon crystal, $U(\textbf{r})$ accounts for the interaction with the impurity ion, $\textbf{E}$ is a uniform external electric field, and $\epsilon$ stands for the resulting energy eigenvalues.

The bottom conduction band in silicon has six equivalent minima (valleys) $\textbf{k}_{0\mu}$, one along each of the crystallographic $\langle 100 \rangle$ directions in $k$ space and located $\approx 86 \%$ of the way to the edge of the Brillouin zone ($\mu=\pm x, \pm y, \pm z$). The donor ground state can be expanded to a good approximation in terms of packets of Bloch functions whose $\textbf{k}$-vectors concentrate around each minimum. This is the cornerstone of EMT \cite{kl1,kl2}, which is improved further by accounting for the inter-valley coupling induced by the impurity potential. Such coupling is strongest in the lattice cell containing the donor nucleus -- this is the so-called central cell correction \cite{ningsah,pantelides,shindonara,hui}. 
In this paper, we will use a multi-valley EMT which accounts for the anisotropy of the silicon conduction band and includes a suitable donor-dependent pseudopotential that mimics the impact of a dopant nucleus on the periodic environment of the undoped silicon lattice hosting the electron \cite{ningsah}.  Within the central cell, the impurity potential differs significantly from the screened Coulomb attraction usually considered in single valley-treatments \cite{kl1,kl2}, and is responsible for the lifting of the valley degeneracy inherent to undoped silicon. We take:
\begin{equation}\label{ning}
U(\textbf{r})=-\frac{e^{2}}{\epsilon_{Si}|\textbf{r}|}(1-\text{e} ^{- b |\textbf{r}|}+ B |\textbf{r}| \text{e} ^{- b |\textbf{r}|}) \equiv -\frac{e^{2}}{\epsilon_{Si}|\textbf{r}|}+U_{cc}(\textbf{r}),
\end{equation}
where $\epsilon_{Si}=11.9$ is the static dielectric constant for silicon, $e$ is the elementary charge, and $b$ and $B$ are parameters setting the two inverse lengthscales specific to the central cell corrections $U_{cc}(\textbf{r})$ of each impurity potential.

After the usual EMT expansion in terms of the Si Bloch functions $\phi_{0}(\textbf{k},\textbf{r})\equiv u_{0}(\textbf{k},\textbf{r}) e^{i \textbf{k}\cdot \textbf{r}}$ with $\textbf{k}$-vector close to each of the six $\textbf{k}_{0\mu}$ \cite{kl1}, we have
\begin{align}\nonumber
\Psi(\textbf{r}) &\equiv \sum_{\mu}\alpha_{\mu} \xi_{\mu}(\textbf{r}) \\
\label{waveexpansion} &= \sum_{\mu}\alpha_{\mu}\frac{1}{(2 \pi)^{3}} \int \tilde{F}_{\mu}(\textbf{k}_{\mu}+\textbf{k}_{0\mu})\phi_{0}(\textbf{k}_{\mu}+\textbf{k}_{0\mu},\textbf{r}) d\textbf{k}_{\mu},
\end{align}  
with $\xi_{\mu}$ being the contribution of the envelope of Bloch functions centered at $\textbf{k}_{0\mu}$. Following the other EMT approximations \cite{hui}, the expectation value of Hamiltonian (\ref{eqbase}) for the wave function (\ref{waveexpansion}) is
\begin{gather}\nonumber
\int d\textbf{r} \sum_p\alpha^{\ast}_{p}F^{\ast}_{p}(\textbf{r}) \times [\alpha_{p} (\textbf{p}\cdot \textbf{A}_{i}\cdot \textbf{p}+e \textbf{E} \cdot \textbf{r}-\epsilon) F_{p}(\textbf{r}) +\\ 
\label{definitiva}\sum_q \alpha_{q} e^{-i(\textbf{k}_{0p}-\textbf{k}_{0q})\cdot \textbf{r}} C_{0}(\textbf{k}_{0q},\textbf{k}_{0p}) U(\textbf{r}) F_{q}(\textbf{r})] =0  ,
\end{gather} 
where the sums over $p$ and $q$ are over the six valley minima; $\textbf{p}\cdot\textbf{A}_{\mu}\cdot\textbf{p}\equiv T$ is the anisotropic kinetic energy operator, which implements the Hamiltonian of the undoped silicon lattice -- first two terms in (\ref{eqbase}) -- through two distinct effective masses ($m^{\ast}_{\perp}=0.191 m_{0}$ and $m^{\ast}_{\parallel}=0.916 m_{0}$), corresponding respectively to perpendicular and parallel motion with respect to each $\hat{\mu}$ axis. $C_{0}(\textbf{k}_{0q},\textbf{k}_{0q}) = 1,  C_{0}(\textbf{k}_{0q},\textbf{k}_{0-q})= -0.1728 $ and $ C_{0}(\textbf{k}_{0q},\textbf{k}_{0\pm p}) = 0.4081 (p\neq q)$, as further detailed in Ref.~\onlinecite{PhysRevB.89.235306}, are due to the lattice-periodic portion of the Bloch functions involved: $u_{0}^{\ast}(\textbf{k},\textbf{r})u_{0}(\textbf{k}',\textbf{r})=\sum_{\textbf{G}}C_{\textbf{G}}(\textbf{k},\textbf{k}') e^{i \textbf{G}\cdot \textbf{r}}$ (where $\textbf{G}$ runs over the vectors of the silicon reciprocal lattice) \cite{kl1}. In particular, EMT requires the Umklapp $\textbf{G}\neq 0$ contributions to this product to be neglected.

\subsection{Zero-field}
With the external electric field turned off, we arbitrarily fix the two donor-dependent parameters of the pseudopotential (\ref{ning}), $b$ and $B$, then variationally minimize the $1s$-manifold energies 
\begin{equation}\label{minimization}
\epsilon_{1s}=\Inf_{\text{\small{ $\Psi^{1s}$}}}\{\langle \Psi^{1s}(\textbf{r})|H|\Psi^{1s}(\textbf{r}) \rangle : \langle \Psi^{1s}(\textbf{r})|\Psi^{1s}(\textbf{r}) \rangle = 1 \},
%\delta\left(\frac{\langle \Psi^{1s}(\textbf{r})|H|\Psi^{1s}(\textbf{r}) \rangle}{\langle \Psi^{1s}(\textbf{r})|\Psi^{1s}(\textbf{r}) \rangle }\right) =0,
\end{equation}
thus setting the corresponding optimal wave functions $\bar{\Psi}^{1s}$. The procedure is repeated with different values for $b$ and $B$, until the experimental ionization energies of the singlet $A_{1}$ (the ground state), the triplet $T_{2}$ and the doublet $E$ eigenstates are reproduced. Those states are the result of the lifting of the six-fold aforementioned valley degeneracy, and their coefficients $\{\alpha_{q}\}$ (see Eq. \ref{waveexpansion}) are fixed by tetrahedral symmetry \cite{kohnbook}, consistent with the pseudopotential employed here. 

Previous multi-valley EMT studies \cite{fritzsche,ningsah,shindonara,hui} have also employed a variational approach, but with hydrogenic Bohr functions as trial effective mass envelopes $F_{q}(\textbf{r})$, to compute the same energy levels. While the variational method is expected to give reliable guesses at the binding energies, no rigorous inferences can be drawn about the exact nature of the wave function: this is the reason why many different pseudopotentials and EMT approximations used in the past led to satisfactory agreement with the former, but poor descriptions of the latter. As a first improvement, in a previous paper \cite{PhysRevB.89.235306} we highlighted the importance of using anisotropic envelopes and imposing further constraints on the shape of the wave function, as indicated by experimental measurements. More precisely, we set the trial ground state function of a Si:P electron to match the experimental contact hyperfine coupling, which is proportional to the value of the electron density at the impurity site. 
%\textcolor{green}{It is better to introduce Table I later, since at this stage the theoretical values of the table are not justified yet.} \sout{We report such parameter matching for all four donors under study here (P, As, Sb and Bi) in Table \ref{tab:hyperfine}.}

\begin{table}  
\begin{center}
  \begin{tabular}{l@{\hspace{10pt}} *{5}{c}}
\hline 
\textbf{Donor} & b (nm$^{-1}$) & B (nm$^{-1}$) & $\epsilon_{A_{1}}^{exp}$ (meV)\cite{ramdas}& $\epsilon_{A_{1}}^{th}$  (meV)   \\
\hline
P & 8.55 & 37.06 & -45.59 & -45.75  \\
\hline
As & 17.74 & 136.84 & -53.76 & -53.54  \\
\hline
Sb & 33.58 & 386.44 & -42.74 & -42.92   \\
\hline
Bi &  48.46 & 1055.7 & -70.98 & -71.08  \\
\hline
\end{tabular}
\caption{Pseudopotential parameters $b$ and $B$ as defined in Eq. \ref{ning} for various V group donors leading to best agreement of the theoretical ground energy $\epsilon_{A_{1}}^{th}$ with its experimental counterpart $\epsilon_{A_{1}}^{exp}$.}
\label{tab:pseudo}
\end{center}
\end{table}
When trying to extend the same approach to include donors other than P, however, we found that matching both binding energies and hyperfine coupling at the same time 
%{\color{red} Giuseppe, what requirements to you mean here?} \textcolor{green}{Added brackets to explain better.} {\color{red} Also, to now it seems like you are doing a multi valley theory, and then suddenly switch to single valley - can you explain this more carefully?} \textcolor{green}{Now I only mention below that the single-valley limit is a good solution far from the nucleus. My discussion of the envelopes here holds for all valleys at the same time. I make now clear that the envelopes of our first paper do not work for Sb and Bi as central cell corrections need a more careful account, i.e. more complicated envelopes.} 
cannot be satisfied for Sb and Bi: more strongly non-isocoric donors (i.e. those that are more different from the hosting silicon atoms) \cite{pantelides} display a larger contact hyperfine coupling, and are expected to need a more careful account of the central cell corrections. Nonetheless, the single-valley limit is a trustworthy solution far enough from the nucleus \cite{kl1}, where the screened Coulomb interaction represents a good approximation to the potential felt by the effective mass electron. For those donors, anisotropic Bohr envelopes are too simple to mediate between those contrasting features.  
%%constraining the electronic density around the impurity site to much larger values than those provided by single-valley treatments,
%%with only two similar lengthscales available (anisotropic Bohr radii), would force too strong deviations from those far-ranged limits. 
We highlight here how more accurate pictures of the electronic spatial density, both close to and far from the nuclear region, can be achieved if envelopes with more structure are used to describe the donor wave function. Two different pairs of anisotropic Bohr radii that distinguish the short ($a_{s},b_{s}$) from the long ($a_{l},b_{l}$) range hydrogen-like decay, and a relative weight $\beta$ of the two parts, define our trial envelopes:
\begin{eqnarray}\label{newtrial}
\begin{aligned}
F^{0}_{z}= N_{0} \left[\text{e}^{-\sqrt{\frac{x^{2}+y^{2}}{a^{2}_{s}}+\frac{z^{2}}{b^{2}_{s}}}} + \beta \hspace{1mm} \text{e}^{-\sqrt{\frac{x^{2}+y^{2}}{a^{2}_{l}}+\frac{z^{2}}{b^{2}_{l}}}}\right], \\
F^{0}_{x}= N_{0} \left[\text{e}^{-\sqrt{\frac{z^{2}+y^{2}}{a^{2}_{s}}+\frac{x^{2}}{b^{2}_{s}}}} + \beta \hspace{1mm} \text{e}^{-\sqrt{\frac{z^{2}+y^{2}}{a^{2}_{l}}+\frac{x^{2}}{b^{2}_{l}}}}\right],
\end{aligned}
\end{eqnarray}    
where $N_{0}$ is a normalization factor. When looking for the optimal solutions in Eq.~(\ref{minimization}), $a_{s},b_{s}$ are essentially fixed by the central cell potential (i.e. they depend strongly on $b$ and $B$), while $\beta$, $a_{l}, b_{l}$ set the resultant long-distance tail, which depends on the screened Coulomb potential surviving further from the nucleus.

This approach is inspired by the observation that, even with only one valley [e.g. setting $\alpha_{1}=1, \alpha_{q}=0$ if $q\neq 1$ in Eq. ~(\ref{definitiva})], the Hamiltonian to be solved is that of a screened hydrogen atom with an extra short range potential, hence the principal quantum number $n$ which labels the radial eigenfunctions of the hydrogen atom is not an exact quantum number for the $s$ states. Anisotropic exponentially decaying shapes are known to provide reliable solutions for a Coulomb-bound electron with two different effective masses along orthogonal spatial directions (see parameters $a_{l}, b_{l}$ above) \cite{kl1}, and the nature of the central cell potential in Eq. (\ref{ning}) suggests the same \emph{ansatz} for the ($a_{s}, b_{s}$) part \cite{ningsah}.

The pseudopotential values that fit $1s$ energies and hyperfine coupling for each donor are reported in Table \ref{tab:pseudo}, alongside the relative ground state energies; expected electronic densities at the nuclear site are listed in Table \ref{tab:hyperfine}, together with the corresponding values deducible from measurements \cite{feher}. The optimal parameters which characterize all ground wave functions, and are used to calculate theoretical values in Tables \ref{tab:pseudo} and \ref{tab:hyperfine}, are listed in Table \ref{tab:parameters}.

\begin{table}
  \begin{tabular}{l@{\hspace{10pt}} *{4}{c}} 
\hline
\textbf{Donor} & A$_{0}$ (MHz) \cite{feher} & $|\Psi(0)|^{2}_{exp} (cm^{-3})$ \cite{feher} & $|\Psi(0)|^{2}_{th} (cm^{-3})$ \\
\hline
P & 117.53 & $0.43\times10^{24}$ & $0.46\times10^{24}$ \\
\hline
As & 198.35 & $1.73\times10^{24}$ & $1.78\times10^{24}$ \\
\hline
Sb & 186.80 & $1.18\times10^{24}$ & $1.15\times10^{24}$ \\
\hline
Bi & 1475.4 & $1.4\times10^{25}$ & $1.4\times10^{25}$ \\
\hline
\end{tabular}
\caption{Theoretical values are calculated as $|\Psi(0)|_{th}^{2}=6 \eta |F^{0}(0)|^{2}$, where $\eta=|u_{0}(\textbf{k}_{0},0)|^{2}/\langle |u_{0}(\textbf{k}_{0},\textbf{r})|^{2}\rangle_{\text{unit cell}} =159.4$ is taken from Ref.~\onlinecite{assali}, and $F^{0}$ is either envelope in Eq. \ref{newtrial}.} 
\label{tab:hyperfine}
\end{table}\
\begin{table*}  
\begin{center}
  \begin{tabular}{l@{\hspace{8mm}} *{6}{c}}
\hline 
\textbf{Donor} & $\bar{a}_{s}$ (nm) & $\bar{b}_{s}$ (nm) & $\bar{\beta}$ & $\bar{a}_{l}$ (nm) & $\bar{b}_{l}$ (nm)  \\
\hline
P & 0.303  & 0.181  & 0.92285 & 1.71 &  0.912 \\
\hline
As & 0.192 & 0.114 & 0.47403 & 1.45 & 0.737  \\
\hline
Sb & 0.146  & 0.0852  & 0.47289 & 1.67 & 0.889   \\
\hline
Bi & 0.0968 & 0.0572 & 0.27153 & 0.967 &  0.472 \\
\hline
\end{tabular}
\caption{Wave function parameters for the donor ground state as defined in Eq. (\ref{newtrial}), found by variational minimization as shown in Eq.~(\ref{minimization}). All long-range radii $\bar{a}_{l}, \bar{b}_{l}$ are significantly smaller than the Kohn-Luttinger values $a_{\rm KL}=2.365$~nm, $b_{\rm KL}=1.36$~nm \cite{kl1}, due to central-cell corrections. Though Si:Sb is more non-isocoric than Si:P, their wave functions look similar in the region far from the nucleus, in line with their similar ground binding energies.}
\label{tab:parameters}
\end{center}
\end{table*}

\subsection{Field on}

The solution of the problem of a hydrogen atom in vacuum within a uniform external electric field $\textbf{E}$ (the Stark effect) has long been known \cite{friedrich}. Perturbation theory correctly predicts, for small fields, quadratic shifts of the ground state energy, linear terms in $|\textbf{E}|\equiv E$ being prevented by parity symmetry. The curvature can only be calculated precisely, though, after an infinite sum over all excited orbital states is admixed into $1s$. 

An alternative approach is supplied by variational theory: the \emph{ansatz} for the ground state under a uniform electric field \cite{pendo} is inspired by the first order perturbative correction to the wave function: 
\begin{equation}\label{hydrogen}
\psi(\textbf{r})=[1+(q_{1}+q_{2} r)z]e^{-r/a_{\rm B}},
\end{equation}  
with $a_{\rm B}$ the Bohr radius, $r=\sqrt{x^{2}+y^{2}+z^{2}}$, while the variational coefficients $q_{1}$ and $q_{2}$ represent the weight of higher orbital states coupled to the fundamental one, and are determined via the principle of minimization of the binding energy of the state in Eq. (\ref{hydrogen}).

The Stark effect in vacuum is complicated, in the framework of shallow donor states in silicon, by two factors: $i)$ as shown in the previous paragraph, the zero-field potential felt by the donor electron, has a short-ranged impurity potential component on top of the (screened) Coulomb interaction. This modifies the response to the field of each separate valley as treated within a single-valley approach, so these are termed {\it intra-valley corrections}); $ii)$ the non-trivial structure of the silicon conduction band introduces the extra valley degree of freedom, hence it becomes important to account for the rearranging of the {\it inter-valley interactions} under $E\neq 0$. Our multi-valley EMT provides one of the most straightforward schemes that can capture the interplay between those two features, and also leads to physical insight. \footnote{We assume in the following, within EMT approximations, that the field dependence of the wave function is entirely ascribed to the envelope part of $\Psi(\textbf{r})$, i.e. we do not include the adjustment of the periodic part of the Bloch functions.}

Following these considerations, our trial zero-field envelopes (\ref{newtrial}) are modified as \cite{PhysRevLett.94.186403}
\begin{eqnarray}\label{newenvelope}
\begin{aligned}
F_{z}= N_{z} \left[\text{e}^{-\sqrt{\frac{x^{2}+y^{2}}{a^{2}_{s}}+\frac{z^{2}}{b^{2}_{s}}}} + \beta \hspace{1mm} \text{e}^{-\sqrt{\frac{x^{2}+y^{2}}{a^{2}_{l,z}}+\frac{z^{2}}{b^{2}_{l,z}}}}\right] (1+ q_{z} z), \\
F_{x}= N_{x} \left[\text{e}^{-\sqrt{\frac{z^{2}+y^{2}}{a^{2}_{s}}+\frac{x^{2}}{b^{2}_{s}}}} + \beta \hspace{1mm} \text{e}^{-\sqrt{\frac{z^{2}+y^{2}}{a^{2}_{l,x}}+\frac{x^{2}}{b^{2}_{l,x}}}}\right] (1+ q_{x} z),
\end{aligned}
\end{eqnarray}
with $q_{x}, q_{z}, a_{l,x}, b_{l,x}, a_{l,z}, b_{l,z}$ being variational parameters, as we justify later on. This procedure is expected to give an appropriate account of the adjustment of each valley to the altered electrostatic environment. Another novelty of our theory is that, for each fixed $\textbf{E}$ value, we choose to minimize the complete singlet $A_{1}$ ground energy, noting that at $\textbf{E}=0$ this is a symmetric superposition of all valleys ($\alpha_{\mu}=1/\sqrt{6}$ $\forall \mu$ in Eq.~\ref{waveexpansion}). Earlier works \cite{PhysRevLett.94.186403,pendo} have optimized the binding energy relative to each valley alone. This is a crucial step forward that allows us to treat valley-orbit effects in a non-perturbative way, and to depict consistently how they are modified by inhomogeneous potentials. 

Let us write the Hamiltonian above in matrix form, in the valley basis $\{\xi_{\mu}\}$. If we assume, with no loss of generality, that $\textbf{E}\parallel \hat{\textbf{z}}$ \cite{PhysRevLett.94.186403}, we have:
\begin{equation} \label{matrix}
H=\left(
\begin{array}{cccccc}
\Lambda_{x} & \Delta_{1x} & \Delta_{2xy} & \Delta_{2xy} & \Delta_{2xz} & \Delta_{2xz} \\
\Delta_{1x} & \Lambda_{x} & \Delta_{2xy} & \Delta_{2xy} & \Delta_{2xz} & \Delta_{2xz} \\
\Delta_{2xy} & \Delta_{2xy} & \Lambda_{x} & \Delta_{1x} & \Delta_{2xz} & \Delta_{2xz} \\
\Delta_{2xy} & \Delta_{2xy} & \Delta_{1x} &  \Lambda_{x} & \Delta_{2xz} & \Delta_{2xz} \\
\Delta_{2xz} & \Delta_{2xz} & \Delta_{2xz} & \Delta_{2xz} & \Lambda_{z} & \Delta_{1z} \\
\Delta_{2xz} & \Delta_{2xz} & \Delta_{2xz} & \Delta_{2xz} & \Delta_{1z} & \Lambda_{z} \\
\end{array}
\right)
\end{equation}
where $H_{\mu \nu}=\langle \xi_{\mu}|H|\xi_{\nu}\rangle$. Diagonal entries $\Lambda_{\mu}$ correspond to intra-valley energies, while $\Delta_{1\mu}, \Delta_{2\mu \nu}$ terms represent couplings between $(\xi_{\mu},\xi_{-\mu})$ and $(\xi_{\mu},\xi_{\nu})$  respectively.
After diagonalization, the ground eigenvector and eigenenergy are simple functions of those matrix elements \cite{PhysRevLett.94.186403}:
\begin{align}
\label{diagonalization}
\epsilon_{g}=&\frac{1}{2}\biggl( \Lambda_{x}+\Lambda_{z}+\Delta_{1x}+\Delta_{1z}+2\Delta_{2xy}\biggr.\\
&\biggl.+\sqrt{32 \Delta_{2xz}^{2}+(\Lambda_{x}-\Lambda_{z}+\Delta_{1x}-\Delta_{1z}+2\Delta_{2xy})^{2}}\biggr)\nonumber,
\end{align}
and
\begin{equation}
\label{alpha}\{\alpha_{\mu}\}_{g}=\frac{(1,1,1,1,\gamma,\gamma)}{\sqrt{4+2 \gamma^{2}}},
\end{equation}
where
\begin{align}
\nonumber
\gamma=&\frac{1}{4 \Delta_{2xz}}\biggl[-(\Lambda_{x}-\Lambda_{z}+\Delta_{1x}-\Delta_{1z}+2\Delta_{2xy})\biggr.\\
&\label{weight}\biggl.+\sqrt{32 \Delta_{2xz}^{2}+(\Lambda_{x}-\Lambda_{z}+\Delta_{1x}-\Delta_{1z}+2\Delta_{2xy})^{2}}\biggr].
\end{align}

It will be appreciated that our matrix is different from that appearing in Friesen's theory of Stark effect \cite{PhysRevLett.94.186403} for four reasons: we consider the whole Bloch functions including the lattice-periodic part, rather than the plane-wave part alone; central cell corrections are implemented in a self-consistent way, fit to experimental electronic properties and crafted to coincide with expected limiting behaviours of the wave function; Eq. (\ref{definitiva}) does not involve spurious inter-valley coupling induced by the kinetic portion of the Hamiltonian, in contrast with Twose's equation \cite{fritzsche}; and, finally, our envelopes are not approximated by their amplitude at the impurity site (a constant), since our $U_{cc}(\textbf{r})$ in Eq. (\ref{ning}) is not a contact potential. 

As the effective local electric field due to $U_{cc}(\textbf{r})$ is always much larger in the region close to the nucleus than the external one due to the field (for the parameter regime considered here), the variational parameters of the donor envelopes in Eq. (\ref{newenvelope}) that pertain to that region are not affected significantly by the field. On the contrary, the long range radii $a_{l}, b_{l}$ and the coefficients $q_{x}, q_{z}$, representing the `squeezing' of the wave function in the $z$ direction, encode all the Stark sensitivity of the ground state $\Psi^{1s}$. Distinct $a_{l}, b_{l}$ are allowed for the envelopes $F_{\pm z}$ in the direction of the field, and for the transverse ones $F_{\pm x}, F_{\pm y}$, as they are expected to adjust differently to the $\textbf{E}\parallel \hat{\textbf{z}}$ perturbation. For each fixed $E$, the optimal values $\bar{q}_{x}$, $\bar{q}_{z}$ and $\bar{a}_{l,x}, \bar{b}_{l,x}, \bar{a}_{l,z}, \bar{b}_{l,z}$, that minimize $\epsilon_{g}$, fix all matrix elements in Eq. (\ref{matrix}), whence $\gamma$ and $\{\alpha_{\mu}\}_{g}$ are consequently determined. 

Deviations of  $\Lambda_{x,z}$ from the zero-field values $\Lambda^{0}_{x,z}$ are seen to be respectively one and two orders of magnitude larger than those of the off-diagonal $\Delta_{2\mu\nu}$ and $\Delta_{1\mu}$ in (\ref{matrix}). These inter-valley terms get negligible alterations from the field directly, since higher Fourier components of a linear Hamiltonian potential $\propto E z$ are not able to couple significantly different valleys: they only change due to the weak squeezing of the envelopes of each separate valley in the $z$ direction. Hence, after expansion of Eq.~(\ref{diagonalization}) up to second order in the field, it is possible to approximate
\begin{align}\label{expansion}
\epsilon_{g}-\epsilon_{g}^{0}&\approx \frac{1}{3} (\Lambda_{x}-\Lambda_{z}), \\
\label{gamma}\gamma&\approx 1-\frac{1}{6}\frac{\Lambda_{x}-\Lambda_{z}}{\Delta_{2xz}}.
\end{align}   

\begin{figure*}
  \centering
    \includegraphics[width=\textwidth]{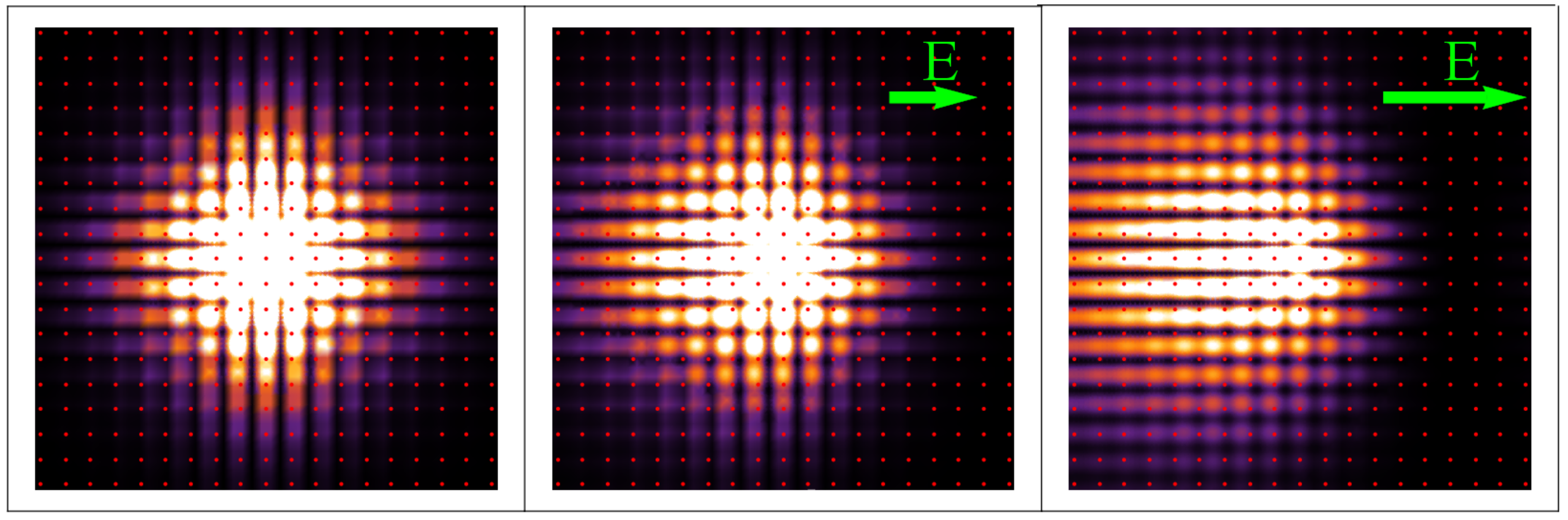}
  \caption{Spatial electronic density of a Si:P bulk donor electron around the implanted nucleus it is bound to, in the plane (010), up to 5~nm away from donor nucleus along the vertical and the horizontal axes. The three panels show how the density changes under different electrostatic environments: The left panel shows the symmetric situation for the isotropic $\textbf{E}=0$ case, the center and the right panel display how the density is driven off the central nucleus in the direction opposite to the vector $\textbf{E}$, under an intermediate and a strong electric field, respectively. Red dots represent the positions of the silicon nuclei of the underlying lattice: their positions do not coincide with the local maxima and minima of the density because of the interference of different valleys contributing to the ground state wave function.}
  \label{densityplot}
\end{figure*}

If we consider the differential equations that lead to the optimal solution in more detail, we can distinguish different trends within the parameter space, as a function of the field. Since $q_{x} a_{l}\hspace{1mm} (q_{z} b_{l})\ll1$ (i.e. the amount of the squeezing of the envelopes in the $z$ direction $\langle F_{x (z)}(\textbf{r})|z|F_{x (z)}(\textbf{r})\rangle$ is very small compared to their zero-field spatial extent), to an excellent degree of approximation 
\begin{align}
\nonumber
\Lambda_{x}\equiv & \hspace{1mm} \Lambda_{x}^{0}+\Delta \Lambda_{x}  \\ 
\approx & \hspace{1mm} \Lambda^{0}_{x}+q_{x}^{2}\langle F^{0}_{x} z |(-\Lambda_{x}^{0}+T+ U(\textbf{r}))|z F^{0}_{x}\rangle \nonumber
\\ & +e E \hspace{1mm} q_{x}\langle F^{0}_{x}| z^{2}|F^{0}_{x}\rangle . & \label{ex}
\end{align}
(The same expressions and discussions presented for the $x$ valleys hold for the $z$ ones, changing $x \rightarrow z$ in the subscripts). Since clearly $\Lambda_{x}^{0}$ does not depend on $q_{x}$, the equation $0=\partial \epsilon_{g}/\partial q_{x}\propto \partial \Lambda_{x}/\partial q_{x}$, which determines $\bar{q}_{x}$, decouples from all others and gives 
\begin{equation}\label{q}
2 \bar{q}_{x}=- \frac{e E \langle F^{0}_{x}| z^{2}|F^{0}_{x}\rangle}{\langle F^{0}_{x} z|(-\Lambda_{x}^{0}+T+ U(\textbf{r}))|z F^{0}_{x}\rangle}.
\end{equation}
As the denominator is positive, it must be that $\bar{q}_{x}, \bar{q}_{z}$ are negative, and their magnitude increases linearly with the field. The wave function extends further along the $z$ axis, in the opposite direction to the vector field $\textbf{E}$ (Fig. \ref{densityplot}). 
This accounts for admixture into the fundamental wave function of $p$ and higher angular momentum orbitals, with the correct singlet $A_{1}$ valley-structure \footnote{The field cannot couple the ground state to other $s$-like orbitals, due to parity symmetry; nor to high angular momentum states which have orthogonal valley structures $T_ {2}$ and $E$.}.

We find, on the other hand, that the radii $a_{l}, b_{l}$ undergo a slight shift from their zero-field values, only to adjust to the new energy terms $\Delta \Lambda_{x,z}$ in Eq. (\ref{ex}). Specifically, the differential equations determining the optimized radii $\bar{a}_{l,x}, \bar{b}_{l,x}, \bar{a}_{l,z}, \bar{b}_{l,z}$ have the form
\begin{align}\nonumber
\epsilon^{0}_{g}[a^{0}_{l}\rightarrow \{a_{l,x}, a_{l,z}\}, a^{0}_{l}\rightarrow \{b_{l,x}, b_{l,z}\}]-\epsilon_{g}^{0} +& \\ \Delta \epsilon_{g} [a^{0}_{l}\rightarrow \{a_{l,x}, a_{l,z}\}, a^{0}_{l}\rightarrow \{b_{l,x}, b_{l,z}\}] &= 0 ,
\end{align}
where $a^{0}_{l}\rightarrow \{a_{l,x}, a_{l,z}\}$ is a shorthand for variations $a^{0}_{l}\rightarrow a_{l,x} (a_{l,z}) $ within each envelope $F_{x} (F_{z})$. The first row is only second order in $\delta a_{l}, \delta b_{l}$ (the unperturbed energy is stationary against small changes of the wave function), while the second includes linear terms in $\delta a_{l}, \delta b_{l}$. Thus, the total shifts of the intra-valley energies $\Lambda_ {\mu}$ [and consequently of the total $\epsilon_{g}$, via Eq. (\ref{expansion})] is due to the parameters $\bar{q}_{x}, \bar{q}_{z}$ alone:
\begin{equation}\label{shift}
\Delta \Lambda_{x,z}\approx e E \langle F_{x,z}(\textbf{r})|z|F_{x,z}(\textbf{r}) \rangle/2 \sim \bar{q}_{x,z} E \sim E^{2}, \\
\end{equation} 
leading immediately, considering Eq. (\ref{expansion}) and (\ref{gamma}), to $\epsilon_{g}-\epsilon_{g}^{0}\propto E^{2}, (\gamma-1) \propto E^{2}$. It then follows that $\{\delta a_{l}, \delta b_{l}\} \propto \{ \bar{q}_{x}, \bar{q}_{z} \} E \propto E^{2}$: these small variations of radii play a decisive role in determining the relative fraction of electronic density leftover at the impurity site, which allows the correct estimates of the Stark shifts presented here.

Let us highlight that, due to the silicon transverse effective mass $m^{\ast}_{\perp}$ being smaller than the longitudinal $m^{\ast}_{\parallel}$, $F_{x}$ extends more broadly in the $z$ direction than $F_{z}$, hence in Eq. (\ref{shift}) $|\Delta \Lambda_{x}|>|\Delta \Lambda_{z}|$: the valleys transverse to the applied perturbation react more effectively than the parallel ones, and this observation will have important consequences, as will become clear when we present our results later on.

\section{Hyperfine Stark shift}

The hyperfine interaction between the electron spin $\textbf{S}$ and the nuclear spin $\textbf{I}$ is described though a coupling tensor $\textbf{A}$
\begin{equation}
H_{\rm HF}=\textbf{I} \cdot \textbf{A} \cdot \textbf{S}.
\end{equation}
The most relevant part of $H_{\rm HF}$, which is usually exploited in quantum computing schemes, is the Fermi contact scalar term $A \hspace{1mm} \textbf{I} \cdot \textbf{S}$ \cite{kane98}, whose values we listed in Table \ref{tab:hyperfine} for all four group V donors. ESR and NMR donor spectra are determined primarily by the interplay between hyperfine and Zeeman splittings, which can result in non-trivial dependence of spin transition frequencies on the background magnetic field $B$, with interesting applications in single qubit control \cite{clocktransitions}. Such features have been exploited in numerous proposals, with successful experimental realizations already achieved in some cases \cite{PhysRevLett.105.067602,newstark,natureactivation}. In particular, the ability to tune these resonant frequencies with external electrostatic gates has often been exploited in proposals \cite{proposal,lansbergen2008}. In the presence of a modified electrostatic environment, the electronic density can be pulled off the impurity site, and thus the hyperfine coupling can be altered. The pertinent regime for  quantum computing schemes, and which will also be explored by measurements reported in the Sec.~\ref{experiments}, is one of weak fields -- of order a few tenths of a V/$\mu$m, which is  well below ionization threshold \cite{kane98}. As stated previously, in this regime we expect a quadratic dependence on $E$: 
\begin{equation}\label{hypshift}
\frac{\Delta A}{A_{0}}\equiv \frac{|\Psi(\textbf{E}\neq 0,\textbf{r}=0)|^{2}}{|\Psi(\textbf{E}= 0,\textbf{r}=0)|^{2}}-1\equiv\eta_{a} E^{2}.
\end{equation}
Unlike $\epsilon_{g}$, the coefficient $\eta_{a}$ is significantly influenced by the precise value of the long range radii $\bar{a}_{l}, \bar{b}_{l}$:
\begin{equation}\label{etaa}
\eta_{a}=\left(\frac{1}{6 F^{0}(\textbf{0})^{2}} \frac{4}{4+2 \gamma^{2}}[2 F_{x}(E,\textbf{0})+\gamma F_{z}(E,\textbf{0})]^{2}-1\right)\frac{1}{E^{2}},
\end{equation}
where $F^{0}(\textbf{0})$ is the value of any zero-field envelope evaluated at the nuclear site. $\gamma$ depends on $E^{2}$, but much more weakly, hence it can be effectively considered equal to $1$ for the evaluations of $\Delta A$ below. 

The optimal parameters enter Eq.~(\ref{etaa}) essentially through the normalizations $N_{x}=\langle F_{x}|F_{x} \rangle, N_{z}=\langle F_{z}|F_{z} \rangle$, since $F_{x (z)}(\textbf{0})=N_{x (z)}(1+\beta)$ [see Eq.~(\ref{newenvelope})]. 
Let us highlight the two ways $\Delta A$ depends quadratically on $E$: From Eqs. (\ref{hypshift}) and (\ref{etaa})
\begin{equation}
\Delta A= \frac{1}{9} \left( 2 \hspace{1mm} \frac{N_{x}}{N_{0}}+\frac{N_{z}}{N_{0}}\right)^{2}-1,
\end{equation}
where $i)$ $N_{x} (N_{z}) \propto q_{x}^{2} (q_{z}^{2}) \propto E^{2}$ (for parity symmetry reasons, $N_{x}$ and $N_{z}$ cannot comprise linear terms in $\bar{q}_{x}, \bar{q}_{z}$, being expectation values of the identity, an even operator); $ii)$ to lowest order, $(N_{x (z)}/N_{0}-1)\approx $  $\frac{1}{N_{0}}\left( \frac{\partial N_{0}}{\partial a_{l}} \delta a_{l,x (z)}+ \frac{\partial N_{0}}{\partial b_{l}} \delta b_{l,x (z)} \right)$, then from the discussion at the end of the previous section we know $\{\delta a_{l}, \delta b_{l}\} \propto E^{2}$.  

\begin{table} 
\begin{tabular}{l@{\hspace{10pt}} *{3}{c}}
\hline
\textbf{Donor} & $\eta_{a} (\mu m^{2}/V^{2})$ (th) & $\eta_{a} (\mu m^{2}/V^{2})$ (exp)\\
\hline
P & -3.0 $\times 10^{-3}$ & $-(2.5 \pm 0.5)\times 10^{-3}$ \\
\hline
As & -1.2$\times10^{-3}$ & $-(1.2 \pm 0.1) \times 10^{-3}$ \\
\hline
Sb & -3.7$\times 10^{-3}$ & $-(3.5 \pm 0.05) \times 10^{-3}$ \\
\hline
Bi & -0.16$ \times 10^{-3} $& $-(0.26 \pm 0.05) \times 10^{-3} $\\
\hline
\end{tabular}
\caption{Quadratic Stark shift coefficients $\eta_{a}$(th) of the hyperfine couplings of four group V donors in silicon, as calculated from Eq. (\ref{etaa}), and compared to respective experimental values $\eta_{a}$(exp) found in Sec.~\ref{experiments}. As we discuss in Sec.~\ref{experiments}, the P, As and Sb donors are measured in the same sample and so here we quote only errors relative to one another; there is an additional absolute error of about 17\% of the shift which is plotted in Fig.~\ref{theory-exp}}
\label{tab:starkhyperfine}
\end{table}    

\begin{figure}[h!]
  \centering
    \includegraphics[width=.49\textwidth]{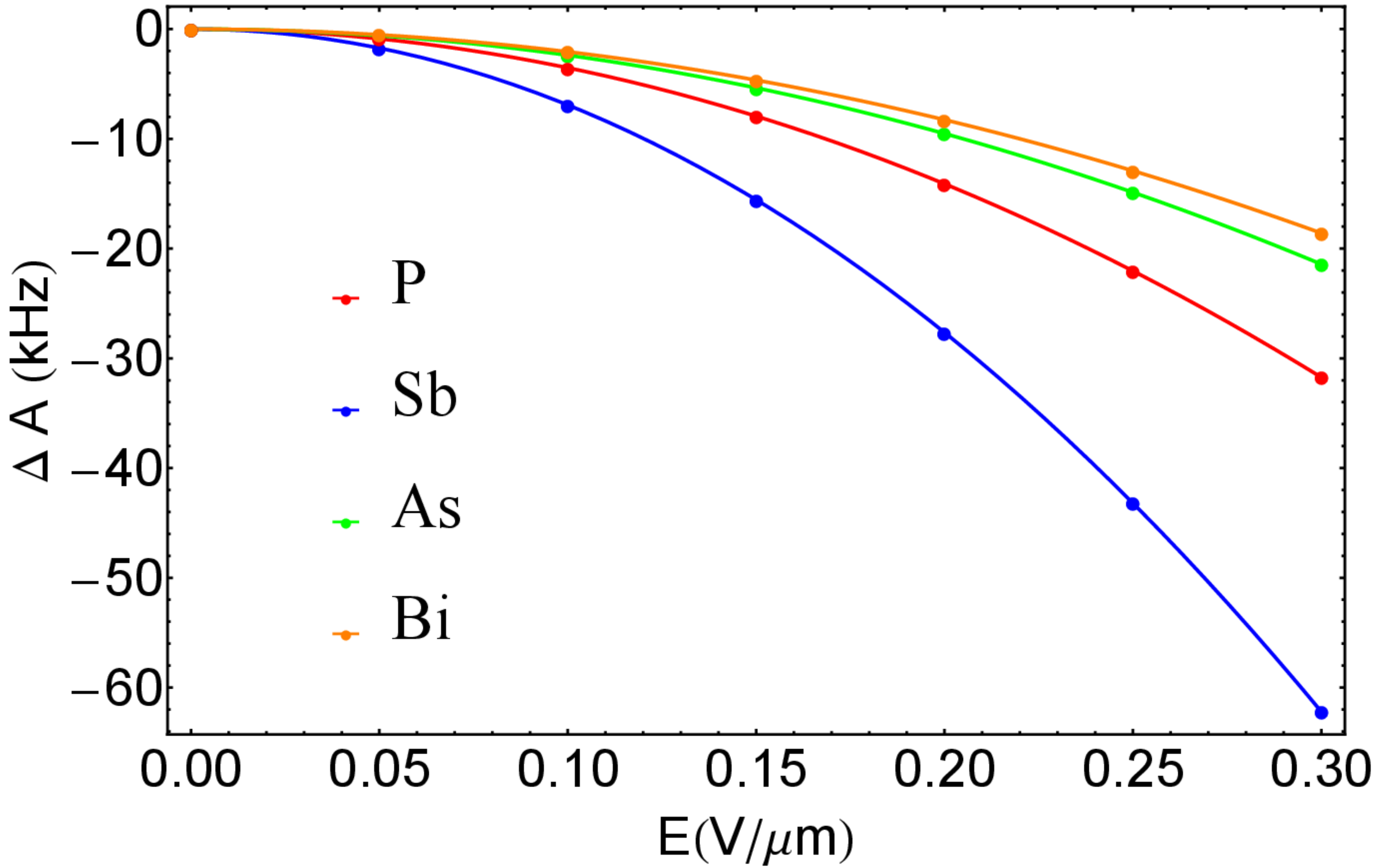}
  \caption{Absolute hyperfine frequency shifts $\Delta A$ as calculated from Eq. (\ref{etaa}), as a function of the applied uniform field $E$, for all donors considered here. The field range shown is typical of those required for executing quantum gates and corresponds to the range investigated experimentally in Sec.~\ref{experiments}, but is well below the ionization thresholds discussed in Sec.~\ref{ionization}.}
  \label{hypgraphs}
\end{figure}

Hyperfine frequency shifts for each donor are displayed as a function of applied field in Fig. \ref{hypgraphs}. From least-squares fitting of those graphs, we obtain values for the quadratic Stark shift coefficient $\eta_{a}$ of hyperfine couplings of all donors considered here; these are shown in Table \ref{tab:starkhyperfine}, alongside their respective experimental values, which have been measured for this study to high precision as will be detailed in Sec.~\ref{experiments}.  

The agreement is excellent for P, As and Sb, and good for Bi; the latter is very non-isocoric \cite{pantelides}, thus an effective mass treatment is expected to work not as well. More specifically, the Umklapp valley-orbit terms neglected in Eq.~(\ref{definitiva}) are more important, and the EMT approximations are less justified. Nonetheless, the Stark shift of Bi is still correctly found to be the lowest of the V group donors.

\begin{figure}[h!]
  \centering
    \includegraphics[width=.49\textwidth]{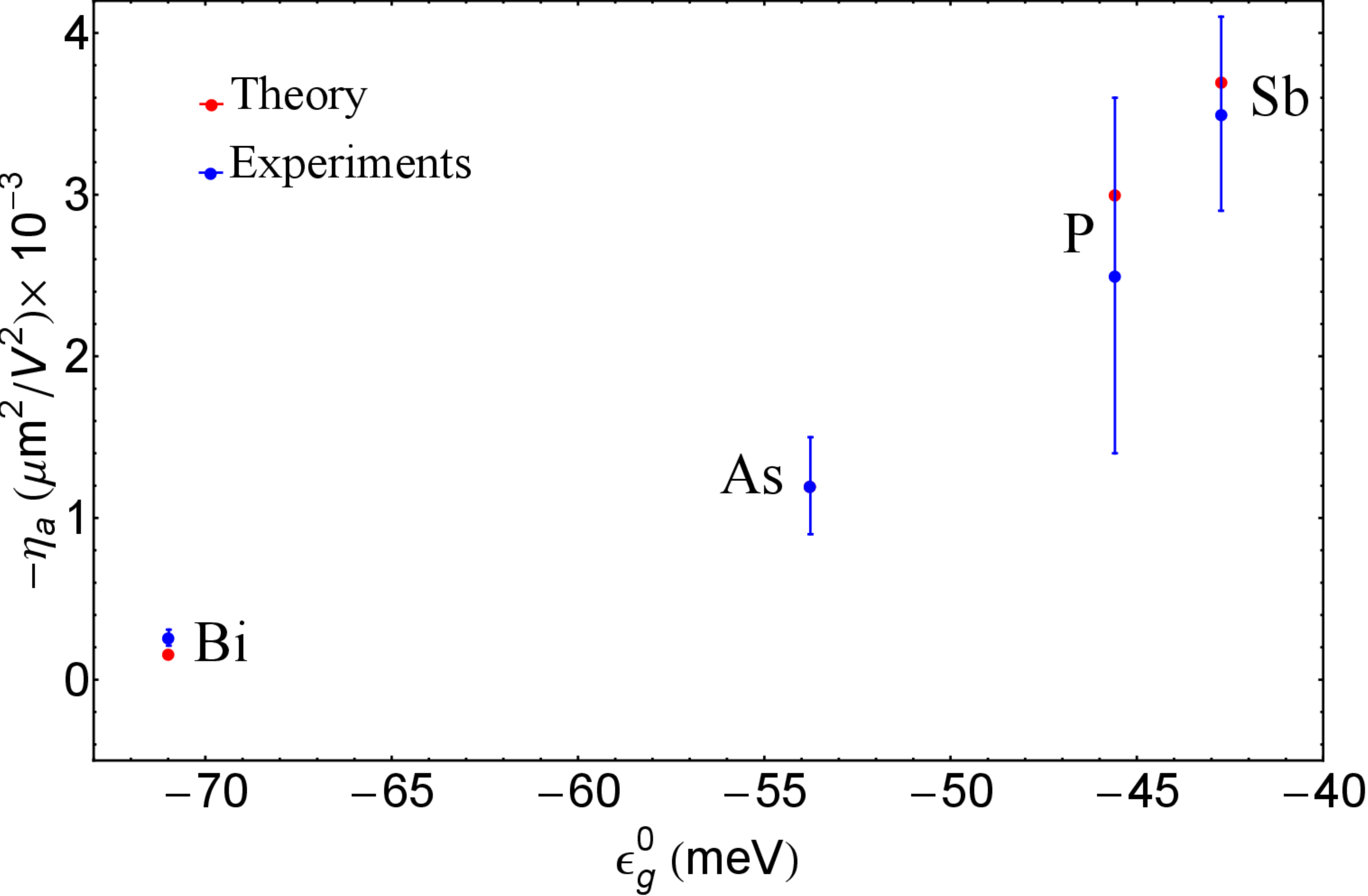}
  \caption{Donor $\eta_{a}$ coefficients as a function of the zero-field ground binding energy $\epsilon_{g}^{0}$: Both theoretical points (red) and experimental values with absolute errors (blue) are reported. The monotonic dependence displayed here is qualitatively explained in the text. The predicted and the measured $\eta_{a}$ coefficient of Si:As overlap with each other.}
  \label{theory-exp}
\end{figure}

Let us stress that the ordering of the magnitudes of $\eta_{a}$ coefficients across different donors, i.e. the trend in the tendency of the corresponding electron to be pulled off the nucleus, follows the pattern suggested by the donors' binding energies, rather than being dictated by the respective hyperfine couplings as one may naively expect. This is shown in Fig. \ref{theory-exp}, where both theoretical and experimental $\eta_{a}$ coefficients are reported, for all donors, in correspondence to their respective ground binding energies. Specifically, Si:Sb shows the largest $\eta_{a}$ since it is the shallowest of all donors, and in spite of the fact that it has a stronger $A_{0}$ than Si:P (see Table II). In other words, it does not only matter how concentrated the electron is at the nuclear site -- more important is how much the ground state is spread further from the impurity,. This can be deduced from Eq.(\ref{q}), (and a similar expression for $\bar{q}_z$) where all the quantities involved are expectation values on the state $|F^{0} z\rangle$, which has vanishing amplitude around $z\approx 0$.  

\section{Experimental methods}
\label{experiments}
Stark shift experiments were performed on ensembles of spins for the four group V donors in two different samples. Material 1 contains $^{31}$P, $^{75}$As and $^{121}$Sb donors ranging in concentration between $10^{14}$ and $10^{15}$~cm$^{-3}$. Material 2 has Bi in concentration of $2\times10^{15}$~cm$^{-3}$. Both materials are isotopically purified silicon-28 float-zone crystals with below 100 ppm isotope concentration of $^{29}$Si and $^{30}$Si. The measurements were realized in a pulse electron spin resonance (ESR) X-band (0.3~T, 9.7~GHz) Bruker spectrometer at temperatures ranging from 5 to 11~K; for each donor, the temperature is adjusted so that the electron spin-lattice relaxation $T_{\rm 1e} \approx 30-40$~ms). Samples of materials 1 and 2 are sandwiched between two metallic plates in between which the voltage is applied, to generate the field in a parallel plate capacitor configuration. 

\begin{figure}[t]%
\includegraphics[width=\columnwidth]{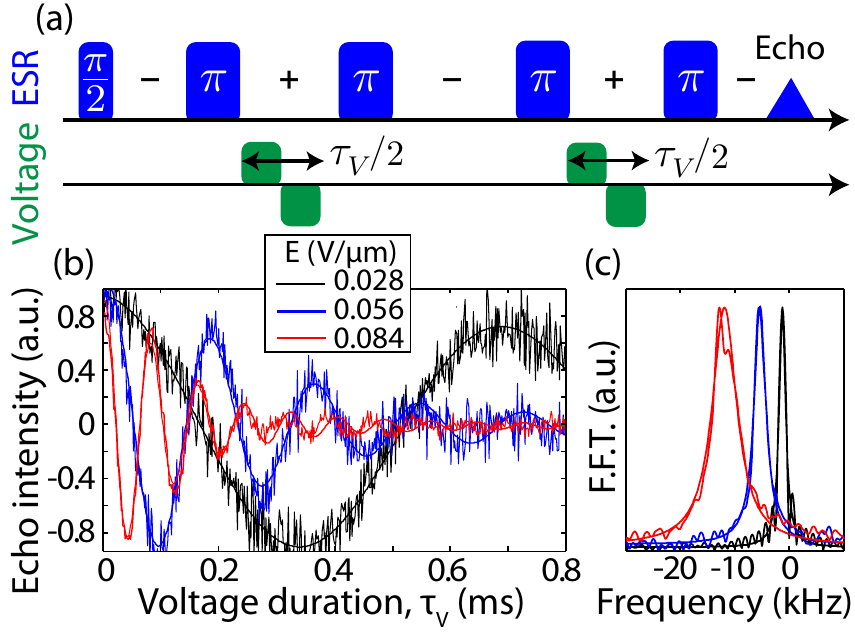}%
\caption{Measurement of the Stark shift in $^{28}$Si:$^{121}$Sb using dynamical decoupling.
(a) Uhrig Dynamical Decoupling (UDD) sequence with four refocusing pulses. As each $\pi$ pulse reverses the phase acquisition (see signs in sequence), the DC electric field is applied in between alternating pairs of $\pi$ pulses. Using positive and negative voltage pulses, the linear Stark shift contribution, arising from local defects, is eliminated and only the quadratic part remains.
(b) Electron spin phase evolution measured in the $m_{\rm I} = +5/2$ ESR transition of Sb in material 1 for different electric fields. (c) Fast Fourier-transform (F.F.T) showing the frequency shift distribution in the sample, fitted here with a Lorentzian fit.
}
\label{sequence}%	
\end{figure}

\begin{figure}[t]%
\includegraphics[width=\columnwidth]{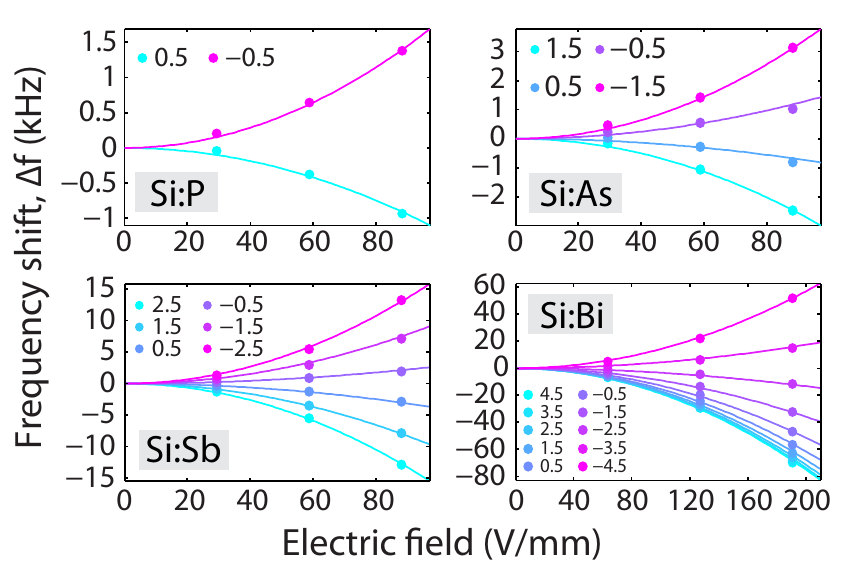}%
\caption{Stark shift for P, As, Sb and Bi measured at X-band, for each possible ESR transitions. The quadratic $\eta_{a}$ and $\eta_{g}$ shift coefficients are the only parameters in the fits (lines). (a) $^{31}$P, (b)$^{75}$As, (c) $^{121}$Sb and (d) $^{209}$Bi.
}
\label{alldonors}%	
\end{figure}

Owing to the Stark effect, the electric field shifts the ESR frequency which can then be measured as a phase shift over time (see Fig.~\ref{sequence}(b)). The frequency shift can be directly retrieved by Fourier-transforming (F.T.) this phase acquisition (see Fig.~\ref{sequence}). Then the frequency shift for all ESR transitions of the four donors can be measured at different electric fields, see Fig.~\ref{alldonors}. The sensitivity of the measurement is thus limited by the frequency deviation and the acquisition time (limited by $T_{\rm 2e}$). 
The electric field deviation, $\sigma_E/\langle E \rangle$, is typically around 15\%, since the plates are not perfectly parallel due to samples geometry and roughness, and the acquisition time can be made as long as 1.6~ms by using a dynamical decoupling sequence (see Fig.~\ref{sequence}(a)). Experimental errors such as variation in sample thickness, voltage pulse rise and set times have also been taken into account. Finally, because P, As and Sb were measured on the same sample of material 1, in exactly the same configuration, the errors in Table~\ref{tab:starkhyperfine} are only given relative to one another, taking into account only the fit error from Fig.~\ref{alldonors}. Due the above mentioned inhomogeneities, there is an additional absolute error of about 17\% of these $\eta_{a}$ values, which were calculated by Monte Carlo sampling over all frequency (F.T.) distributions. This additional error is not included in Table~\ref{tab:starkhyperfine} but is shown in Fig.~\ref{theory-exp}.

Local strain or charge defects in the sample create an internal electric field $E_{\rm in}$. In the presence of an external electric field $E_{\rm ex}$, the Stark shift is:
\begin{equation}
\Delta f\propto(E_{\rm ex}+E_{\rm in})^2=2E_{\rm in}E_{\rm ex}+E_{\rm in}^2 + E_{\rm ex}^2 .
\end{equation} 
As a result, the defect induced Stark shift has a component that depends linearly on the external electric field \cite{PhysRevLett.97.176404}. This component is expected to be strictly inhomogeneous and thus results in a decay of the electron spin echo signal - but it can be cancelled by applying bipolar (positive and negative) electric pulses, as given by the sequence depicted on Fig.~\ref{sequence}(a). The quadratic shift which we want to measure is still acquired under this sequence.

The Stark-induced frequency ($f$) shift combines both the hyperfine ($A$) and the spin-orbit ($g$) contributions. They both depend quadratically on the applied electric field, but the sensitivity of the frequency to each of them ($df/dA$ and $df/dg$, respectively) depends on both the nuclear state $m_I$ and the magnetic field $B_0$. In the high field limit (no mixing) and for the electron spin transition, $df/dA = m_I$ and $df/dg = \mu_B B_0$. For $B_0 < 1$~T, the hyperfine contribution is then expected to be strongly dominant over the spin-orbit contribution. Thus, at X-band, measuring the Stark shift for each of the $m_I$ states provides a good estimate of the hyperfine contribution in very good agreement with theoretical values (see Fig.~\ref{theory-exp}).
The values of $\eta_{a}$ measured here agree well with previous results for As and Sb \cite{arsenicstark,PhysRevLett.97.176404}, while this is the first time they have been reported for P and Bi.

\section{Ground state energy and electron ionization}
\label{ionization}

Using our previous analysis, captured by Eq. (\ref{expansion}), and keeping in mind that $\Lambda_{x}<\Lambda_{z}<0$, we find that the absolute magnitude of the donor binding energy $|\epsilon_{g}|$ increases with increasing field. This is non-trivial behaviour that arises from the combination of effects discussed in Sec. II,
%{\color{red} where does this obsveration come from? It it a bit out of the blue - can you either explain where is comes from here or refer to a previous section?} \textcolor{green}{Better now?} 
where we saw how the lowering of intra-valley energies produced by the field, see Eq. (\ref{shift}), is the only important factor in determining the change in the total donor binding energy with $E$. The impact on the shape of the wave function in the central-cell region modifies $\epsilon_{g}$ negligibly, hence the inter-valley interactions are not affected significantly, and the spectral narrowing of the $1s$ manifold is not strong enough to produce an overall energy increase of the ground state. Our conclusion contrasts with Ref. \onlinecite{PhysRevLett.94.186403}, where the electron was predicted to be less bound in increasing field, but it confirms the conclusions in the \emph{ab initio} treatments presented in Refs.~\onlinecite{Debernardi2006,PhysRevB.80.165314}. 

%Let us highlight that such behavior is also indirectly corroborated by the $g$-factor predictions reported here and in Ref. \onlinecite{PhysRevB.80.155301}: starting from Eq.~(\ref{gamma}), since $\Delta_{2xz}<0$, $\epsilon_{g}-\epsilon_{g}^{0}$ and $\gamma-1$ must have the same sign, hence if for $\textbf{E}\parallel \textbf{B}$ [$\theta=0$ in Eq.~(\ref{gformula})] the $g$-factor decreases with increasing field, then so should $\epsilon_{g}$.

\begin{figure}[h!]
  \centering
    \includegraphics[width=.48\textwidth]{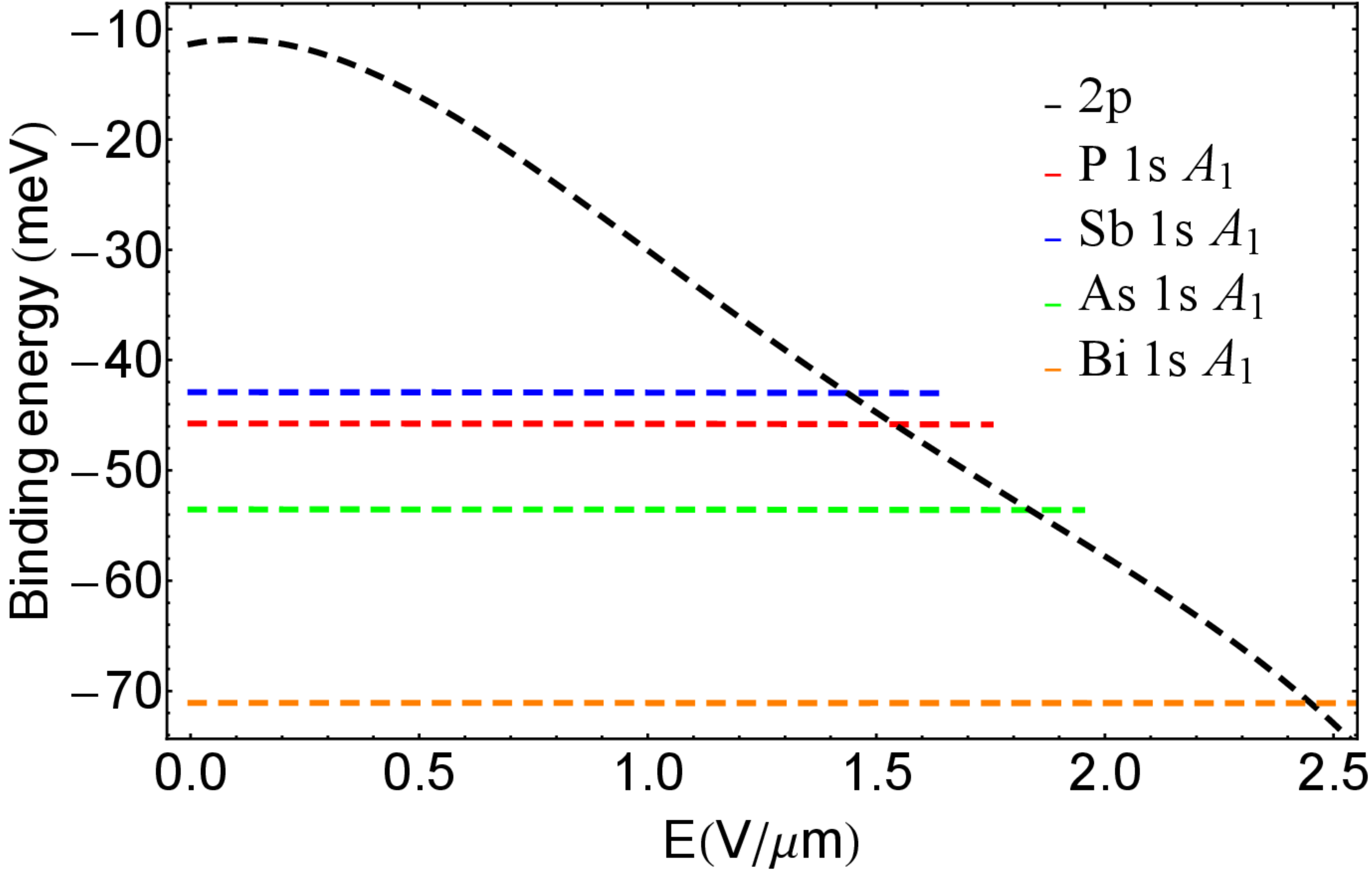}
  \caption{Donor binding energies of the ground states of Si:P, Si:Sb, Si:As, and Si:Bi decrease very weakly as a function of the field, as detailed in the text. For each species, values are reported only up to the point $E_{c}$ where they become degenerate with the `$2p$' energy level, which is common to all donors, as it is not influenced by central cell corrections. Each crossing point corresponds to the ionizing field for each donor.}

  \label{ionizations}
\end{figure}

We find the dependence of the ground state energy on the field is rather weak, as shown in the $1s$ $A_{1}$ energy plots of Fig. \ref{ionizations}: this is compatible with studies performed within different approaches \cite{PhysRevB.80.165314}, and confirms that bulk donor electrons `instantaneously' tunnel off an impurity nucleus, in contrast to adiabatic tuning available to electrons closer to an interface \cite{tb}.

So far we have dealt with electric fields of magnitudes that would be required for the execution of quantum gate operations. As $E$ increases by one order of magnitude above those considered so far, qualitatively new dynamics takes place \cite{Debernardi2006}: the $2 p$-like orbital levels of the donor electron (with the singlet $A_{1}$ valley structure) anti-cross with the slowly changing ground $1s$-like state, so that the electron can effectively tunnel off the bulk of the silicon layer. Using our model we can predict the ionization field for each of the donor chemical species. The size of this field is important, since spin dependent tunnelling is a leading proposed read-out technique for solid state spins \cite{spinmem, proposal}. Such read-out often occurs at the interface with an oxide layer, or close to an SET device \cite{pla12, pla13}, which is rather far from the dopant nucleus and thus requires ionization of the donor electron.

We compute the Stark shifted binding energy of the bulk $A_{1}$ `$2 p_{0}$' state as a function of the field (see Fig. \ref{ionizations}), by variational optimization of Hamiltonian (\ref{eqbase}) on the following trial wave function: 
\begin{equation}
\Psi^{2 p}(\textbf{r})=N_{p} \hspace{1mm} z \hspace{1mm} \text{e}^{-\sqrt{\frac{x^{2}+y^{2}}{a^{2}_{p}}+\frac{z^{2}}{b^{2}_{p}}}} (1+q^{p}_{z} z),
\end{equation}      
which is suggested by the zero-field form in Ref.~\onlinecite{kl2}, modified to include the admixing with higher energy states induced by the applied external field. Let us remark that valley-orbit effects play practically no role in determining the energy and the wave function of this state (and more generally, of all non-$s$ states), since the corresponding orbital is concentrated far from the impurity nucleus and is thus not sensitive to the non-Coulombic potential $U_{cc}(\textbf{r})$. For the same reason, its features do not depend on the specific chemical donor species. 

We stress that the energy levels shown in Fig. \ref{ionizations} refer to the diagonal Hamiltonian terms $\langle \Psi^{2p}|H|\Psi^{2p} \rangle$ and $\langle \Psi^{1s}|H|\Psi^{1s} \rangle$, i.e. we do not take into account the off-diagonal couplings $\langle \Psi^{2p}|H|\Psi^{1s} \rangle$. The latter would lead to the expected anti-crossing of the levels as hybridization between $\Psi^{1s}$ and $\Psi^{2p}$ occurs. However, the field at which $\langle \Psi^{2p}|H|\Psi^{2p} \rangle=\langle \Psi^{1s}|H|\Psi^{1s} \rangle$ provides a good estimate of the ionization field $E_{c}$\cite{externalexchange}.

The dependence of the $A_{1}$ $2p$ binding energy on field qualitatively confirms the behavior calculated in Refs.~\onlinecite{tb, Debernardi2006} and \onlinecite{PhysRevB.80.165314} for a donor electron state closer than 25~nm to the interface with a dioxide. Our results are specific to impurities implanted deep in the bulk of a Si layer, though, and hence there are quantitative differences of a few tenths of V/$\mu$m between our results and the threshold for Si:P and Si:As predicted in those references. We report in Table \ref{tab:ionization}, for the first time, the expected ionization fields for all bulk donors. 

\begin{table}
\begin{tabular}{l@{\hspace{10pt}}c@{\hspace{7pt}}c@{\hspace{7pt}}c@{\hspace{7pt}}c@{\hspace{7pt}}}
\hline
\textbf{Donor} & Ionization   & Maximum & Maximum ESR \\
& field &  hyperfine shift  &  frequency shift \\
&  $E_{c}$ (V/$\mu$m) & $\Delta A^{max}$ (MHz) &  $\Delta f^{max}$ (MHz) \\
\hline
P &  $1.55$ & $0.8$ & $0.4$ \\
\hline
As & $1.84$ & $0.8$ & $1.2$ \\
\hline
Sb & $1.45$ & $1.4$ & $3.5$ \\
\hline
Bi & $2.45$ & $1.4$ $(2.1)$ & $6.3$ $(9.5)$ \\
\hline
\end{tabular}\\
\caption{Predictions of the size of the electric field required to ionize each donor species, and the corresponding maximum absolute hyperfine shift $\Delta A^{max}$ that can be achieved before the electron tunnels away from the nucleus. Each ESR frequency shift $\Delta f^{max}= \Delta A^{max} m_{I}$ in the last column are for a nuclear magnetic moment $m_{I}=I$ (i.e. the maximum possible value of $m_{I}$) and represents the largest transition frequency shift that can be induced by the applied field $E$ with each donor. The first $\Delta A^{max}, \Delta f^{max}$ for Bi are calculated using the theoretical value for $\eta_{a}$ given in Table \ref{tab:starkhyperfine}, while the bracketed value refers to the experimental $\eta_{a}$ measured here and available in Table~\ref{tab:starkhyperfine}.}
\label{tab:ionization}
\end{table}

While specific measurements of these thresholds are still lacking, very recent experimental work \cite{arsenicstark} reports that Si:As donor electrons are ionized at $E\sim 2$ V/$\mu$m, in full agreement with our prediction.

Other than identifying precise field regimes that are relevant for bulk donor spin read-out, our study allows us to extract another piece of information valuable to any silicon quantum computing scheme. Single qubit operations in this system are performed via selective microwave (ESR) magnetic pulses addressing the hyperfine- and Zeeman-split transitions of the donor electron spin levels (in the high magnetic field limit, electron-spin levels are only weakly hyperfine-mixed with the nuclear spin ones, i.e. the electron spin projection $m_{S}$ is a good quantum number). In order to manipulate individual spins within a large ensemble of implanted donors it is easiest\cite{embracing} to apply a global alternating magnetic field $B_{ac}$, bringing only selected qubits in resonance with it, by locally Stark-shifting their spin-resonance frequency \cite{kane98}. The selected ESR transitions can be shifted by at most $\Delta f (E)=\eta_{a} E^{2} A_{0} m_{I}$ with $m_{I}$, the nuclear spin projection, equal to the nuclear spin quantum number $I$ \cite{newstark}. 
This maximum shift sets the limit on how quickly spins can be manipulated: if the timescale $\tau$ of $B_{ac}$ pulses is shorter than $\Delta f^{-1}$, then the resonance frequencies of the non-selected qubits will lie within the pulse bandwidth.
It follows that faster gates can be performed with larger $\Delta f$, and this is in turn limited by the ionization threshold presented here.

We estimate the maximum hyperfine frequency shifts that donor ESR transitions can undergo in silicon while still being safe from ionization: results are reported in Table \ref{tab:ionization}. Si:Bi supports gate times as short as $\Delta f_{max}^{-1}\sim 100$ ns, yielding the fastest manipulation obtainable with Kane-like $A$ gates \cite{kane98} within donor spins systems in silicon.

\section{Conclusions}

Our theory provides the first comprehensive treatment of Stark effects for donors in silicon. The inherent physical mechanisms behind them are unveiled by the analytic and insightful multi-valley EMT framework. After appropriate calibration using bulk donor properties, we obtain an excellent match with experimental hyperfine shifts of all V group donors under a non-zero applied electric field. The reported measurements of hyperfine Stark shifts include the first experimental hyperfine $\eta_{a}$ coefficient of Si:P and Si:Bi. 

We establish that the donor electron is slightly more bound to the nucleus with an increasing field, for small fields, and calculate field thresholds at which ionization is expected to occur, for each donor. This leads us to estimating the maximum frequency shifts of ESR transitions that can be achieved by $A$ gates in a Kane-like architecture. Very short operation times, as fast as $\sim 100$ ns, are allowed if the qubit is implemented in the Si:Bi electron spin.

Building on these results, our reliable wave functions are ready to be used for calculation of other single and two donor electron properties, especially those relevant for implementing quantum information processing protocols. They represent a fast and flexible scheme rich in physical insight, easily extendable to include more complicated electromagnetic environments, such as interfaces, non-uniform electric fields, and hybrid donor-dot schemes.

\section{Acknowledgements}
We thank C.C. Lo and A.M. Tyryshkin for valuable discussions. 
This research was funded by the joint EPSRC (EP/I035536) / NSF (DMR-1107606) Materials World Network grant (BWL, GP, JJLM, SAL), EPSRC grant EP/K025562/1 (BWL and JJLM), the European Research Council under the European Community'��s Seventh Framework Programme (FP7/2007-2013) / ERC grant agreement no. 279781 (JJLM), partly by the NSF MRSEC grant DMR-0819860 (SAL), the Department of Energy, Office of Basic Energy Sciences grant DE-SC0002140 (RNB). BWL and JJLM thank the Royal Society for a University Research Fellowship. RNB thanks the School of Natural Sciences, Institute for Advanced Study, Princeton for hospitality during the period when this work was done. The 28-Si-enriched samples used in this study were prepared from Avo28 material produced by the International Avogadro Coordination (IAC) Project (2004-2011) in cooperation among the BIPM, the IN-RIM (Italy), the IRMM (EU), the NMIA (Australia), the NMIJ (Japan), the NPL (UK), and the PTB (Germany).

%%\bibliographystyle{unsrt}
%%\bibliographystyle{natbib}
%%\bibliography{referencesGP}

\end{document}